\documentclass[12pt]{iopart}
\usepackage{iopams}  

\newcommand{\be}{\begin{equation}}
\newcommand{\ee}{\end{equation}}
\newcommand{\bea}{\begin{eqnarray}}
\newcommand{\eea}{\end{eqnarray}}

\newcommand{\iGAj}{{j}}

\newcommand{\cliffconj}[1] { \bar{#1} }

 \newtheorem{defn}{Definition}

\begin{document}

%
%
%
%
%
%
%
%

\title[Deriving Minkowski spacetime]{The origin of Minkowski spacetime}


\author[Chappell]{James M.~Chappell}
\address{School of Electrical and Electronic Engineering, \\ University of Adelaide, SA 5005 \\ Australia}
\ead{james.chappell@adelaide.edu.au}

\author{John G.~Hartnett}
\address{School of Physical Sciences, \\ University of Adelaide, Adelaide SA 5005 \\ Australia}

\author{Nicolangelo Iannella}
\address{Faculty of Mathematics and Natural Sciences \\ University of Oslo \\ PostBoks 1066,
Blindern 0316, Norway }

\author{Azhar Iqbal}
\address{School of Electrical and Electronic Engineering, \\ University of Adelaide, SA 5005 \\ Australia}

\author{David L. Berkahn}
\address{School of Electrical and Electronic Engineering, \\ University of Adelaide, SA 5005 \\ Australia}

\author{Derek Abbott}
\address{School of Electrical and Electronic Engineering, \\ University of Adelaide, SA 5005 \\ Australia}



\vspace{2pc}


\newpage

\date{November 5, 2015}

\begin{abstract}
The four dimensional spacetime continuum, as first conceived by Minkowski, has become the dominant framework within which to describe physical laws. 
In this paper, we show how this four-dimensional structure is a natural property of physical three-dimensional space, if modeled with Clifford geometric algebra $ C\ell(\Re^3) $.   We find that Minkowski spacetime can be embedded within a larger eight dimensional structure. This then allows a generalisation of the invariant interval and the Lorentz transformations. Also, with this geometric oriented approach the fixed speed of light, the laws of special relativity and  a generalised form of Maxwell's equations, arise naturally from the intrinsic properties of the algebra without recourse to physical arguments.  We also find new insights into the nature of time, which can be described as two-dimensional. Some philosophical implications of this approach as it relates to the foundations of physical theories are also discussed.
\end{abstract}

\maketitle

\section{\label{sec:level1}Introduction}

Einstein's seminal paper of 1905~\cite{Einstein1905} inaugurated a new understanding of space and time, which included the properties of time dilation and length contraction. Minkowski subsequently showed that these exotic phenomena could be  interpreted as simply the intrinsic properties of a four-dimensional spacetime continuum, with an invariant mixed metric  distance $  d s^2 = dt^2 - d \boldsymbol{x}^2 $. It is found that the Lorentz transformations will hold this interval invariant and produce the various spacetime effects.  Lorentz invariance, which has now been extremely well experimentally verified, has become a default requirement for all valid physical laws.  The foundational nature of the Minkowski metric and Lorentz invariance, hence motivates the work in this paper, to more fully understand its origin and possible generalisations.  Note, that we are not attempting a generalisation of Minkowski spacetime to include gravity. However, Minkowski spacetime retains validity within general relativity (GR), as the curved spacetime manifold is asymptotically Minkowskian at each point. Also, the Minkowski metric is actually a solution of Einstein’s field equations, describing the metric for gravity free spacetime, which is generalised as the Friedmann–Lema\^{i}tre–Robertson–Walker (FLRW) metric. 
To give a stronger physical basis, it has been proposed by various authors~\cite{Verlinde2017} that Minkowski spacetime, as well as gravity, arises as an effective description of microscopic degrees of freedom, as in emergent gravity, loop quantum gravity or string theory, for example. It might also be possible to derive the Lorentz invariant structure directly from the properties of the quantum vacuum. This paper also supports an underlying physical basis to Minkowski spacetime, as a result of the interaction of the four distinct physical elements, represented by scalars, vectors, pseudovectors and pseudoscalars. Nevertheless, it is reasonable to assume that Minkowski spacetime will ultimately break down at the Planck scale.


As a way to motivate our approach, one of the most fundamental aspects of the physical world is the existence of three degrees of spatial freedom.
Indeed, the three dimensions of space are assumed by default in Minkowski spacetime.  This idea is indeed supported by the observation of precisely five regular solids~\cite{coxeter1973regular,Oziewicz2012}, which only occurs with exactly three spatial dimensions, as well as the inverse square laws of gravity and electromagnetism, which have been experimentally verified to high precision~\cite{hoyle2001submillimeter}.  
Therefore, it is natural to adopt the formalism of Clifford geometric algebra (GA) of three dimensions $ C\ell(\Re^3) $ in order to describe spacetime.  Previous examples of approaches using GA, or Clifford algebra more generally~\cite{pavsic2006}, include: (i) spacetime algebra (STA)~\cite{Hestenes:1966} and (ii) the algebra of physical space (APS)~\cite{Baylis1996}. Here, STA posits four basis vectors to represent three spatial and one time dimension.  On the other hand, APS works solely with three spatial basis vectors, and represents time separately as a scalar variable.  Both of these approaches can successfully reproduce the results of special Relativity (SR).
The distinction between these approaches, is that the Minkowski mixed metric is assumed by default, whereas our approach attempts to derive this from more fundamental principles.

With our approach, we adopt as a foundational postulate that three-dimensional physical space is represented by the Clifford algebra $ C\ell(\Re^3) $.  From this point of view, we find that time and the Minkowski metric are naturally emergent properties. While recovering standard results as special cases, we  find that spacetime generalises to eight dimensions. Also, while recovering the Lorentz transformations, we find a more general class of transformations, allowing us to predict new kinematical effects.

\section{Deriving Minkowski spacetime from $ C\ell(\Re^3) $}

We begin with our foundational hypothesis, that spacetime is represented by the Clifford algebra $ C\ell(\Re^3) $, being an eight-dimensional graded algebra
$  \Re \oplus \Re^3  \oplus  \bigwedge^2 \Re^3 \oplus \bigwedge^3 \Re^3  $.
To clarify the notation, $ C\ell(\Re^3) $ can be expressed using standard three-vectors as a multivector
\be \label{multivector3DSimple}
M = a + \boldsymbol{x} + \iGAj \boldsymbol{n} + \iGAj b ,
\ee
where $ \boldsymbol{x} = x_1 e_1 + x_2 e_2 + x_3 e_3 $ a vector, $ \iGAj \boldsymbol{n} = n_1 e_2 e_3 + n_2 e_3 e_1 + n_3 e_1 e_2 $ a bivector, where $ \boldsymbol{n} = n_1 e_1 + n_2 e_2 + n_3 e_3 $, and $ \iGAj = e_1 e_2 e_3  $ trivector, with $ a, b, x_1, x_2, x_3, n_1, n_2, n_3 $  real scalars~\cite{Chappell2015,Chappell2014IEEE}.
We have used as a basis for three dimensional Clifford algebra $ C\ell \left ( \Re^3 \right ) $, the three unit elements $ e_1, e_2, e_3 $, having a unit square $ e_1^2 = e_2^2 = e_3^2 = 1 $, and are anticommuting, with $ e_1 e_2 = - e_2 e_1 $, $ e_1 e_3 = - e_3 e_1 $ and $ e_2 e_3 = - e_3 e_2 $. 
The four types of elements within $ C\ell \left ( \Re^3 \right ) $ can be used to describe four distinct physical quantities typically represented as scalars, polar vectors, axial vectors (or pseudovectors) and pseudoscalars, respectively. 
The eight-dimensional multivector in Eq.~(\ref{multivector3DSimple}) is  isomorphic to the complexified quaternions~\cite{hamilton1853lectures,Silberstein1912} and closely related to the octonions~\cite{dray2015geometry,gogberashvili2005octonionic,gogberashvili2016octonionic,gogberashvili2015geometrical,gogberashvili2014split}, which have also been used previously to model spacetime.

\subsection{Key properties of geometric algebra}

The first key property of GA is that no new products need to be defined, such as the dot and cross products, as all products revert to the foundational process of the distribution of multiplication over addition, that is, the standard process of expanding brackets, refer Appendix~A.  
Secondly, the three basis vectors in $ C\ell(\Re^3) $ can produce the bivectors $ e_1 e_2 $, $e_1 e_3 $, $ e_2 e_3 $ and a trivector $ e_1 e_2 e_3 $.  The bivectors naturally represent such physical phenomena as magnetic fields, allowing a unified electromagnetic field variable $ \boldsymbol{E} + \iGAj \boldsymbol{B} $, for example.
Thirdly, the trivector quantity $ \iGAj = e_1 e_2 e_3 $, is isomorphic to the unit imaginary $ i $, thus allowing us to naturally incorporate complex valued functions.
Further, as an aid to visualisation, the four algebraic elements of scalars, vectors, bivector and trivectors represent the four common geometrical entities of points, lines, areas and volumes, respectively.  

\subsection{Finding the invariants}

Due to the close connections between invariants and the laws of Nature, we now explore the invariants in  the space of $ C\ell(\Re^3) $ described by Eq.~(\ref{multivector3DSimple}), after applying the most general transformation rules.

\begin{defn}[Clifford conjugation]
We define {\it Clifford conjugation} of a multivector $ M $  as
\be \label{CliffordConjugation}
\cliffconj{M} =  a - \boldsymbol{x} - \iGAj \boldsymbol{n} + \iGAj b .
\ee
Clifford conjugation is an involution that is an anti-automorphism, so that for a product $ M N $ of two multivectors $ M , N \in  C\ell \left ( \Re^3 \right ) $, $ {\overline{M N}} = \cliffconj{N} \cliffconj{M} $.
\end{defn}

\begin{defn}[Multivector amplitude]
We define the {\it amplitude squared} of a multivector $ M $ through Clifford conjugation, giving the bilinear form
\be \label{AmplitudeSquared}
| M |^2 = M \cliffconj{M}  = a^2 - \boldsymbol{x}^2 + \boldsymbol{n}^2 - b^2 + 2 \iGAj \left (a b - \boldsymbol{x} \cdot \boldsymbol{n}  \right )
\ee
forming a complex-like number $ \in \mathbb{C} $, and thus commuting with the rest of the algebra.  
\end{defn}

We refer to this as a ``complex-like'' number because, as already noted, the trivector $ \iGAj $ is analogous to the unit imaginary and all other quantities are real scalars. The square root is therefore well defined from complex number theory and so we can define the multivector {\it amplitude} as $ | M | = \sqrt{| M |^2 } $.  This is essentially a norm, although in our case, it may be complex. The amplitude squared has the property 
\be
| M_1 M_2 |^2 = M_1 M_2 \cliffconj{M_2} \cliffconj{M_1} = M_1 \cliffconj{M_1} M_2 \cliffconj{M_2}  = | M_1|^2 |M_2|^2 .
\ee
We can therefore write a norm relation
\be
| M_1 M_2 |  = | M_1| |M_2| ,
\ee
provided that the appropriate branch is used when finding the complex square roots.

\subsection{The group transformations}

\begin{defn}[Bilinear multivector transformation]
We define a general bilinear transformation on a multivector $ M $ as
\be \label{generalTransformation}
M' = K M L ,
\ee
where $ M, K, L \in C\ell(\Re^3) $.
\end{defn}
We then find the transformed multivector amplitude
\be \label{invarianceMultivectorTransformation}
| M' |^2 = K M L \, {\overline{K M L}} = K M L \cliffconj{L} \cliffconj{M} \cliffconj{K} = |K|^2 |L|^2 | M |^2 ,
\ee
where we have used the anti-involution property of Clifford conjugation and the important commuting property of the amplitude. Hence, provided we specify a unitary condition $ |K|^2 |L|^2 = 1 $ for these transformations, then the amplitude $ | M | $ will be invariant. Without loss of generality, it is then convenient to impose the condition $ |K |^2 =  | L |^2 = \pm 1 $.  The transformation in Eq.~(\ref{generalTransformation}) is then the most general bilinear transformation that preserves the multivector amplitude and so produces an invariant distance over the space.  The selection of the involution of Clifford conjugation is not arbitrary, as it is the only involution producing a commuting complex-like number allowing these invariants to form, according to Eq.~(\ref{generalTransformation}). 

If we focus on the special case $ |K |^2 =  | L |^2 = +1 $ that describes transformations that are continuous with the identity.
Then, using the power series expansion of the exponential function, the multivector $ K $ can be written in an exponential form~\cite{Hestenes:1966,Hestenes2003}
\be
K = \rme^{ c + \boldsymbol{p} + \iGAj \boldsymbol{q} + \iGAj d } ,
\ee
provided $ |K | \ne 0 $, where $ c, d \in \Re $ and $ \boldsymbol{p},\boldsymbol{q} \in \Re^3 $.
Therefore, we find
\be
K \cliffconj{K} = \rme^{ c + \boldsymbol{p} + \iGAj \boldsymbol{q} + \iGAj d } \rme^{ c - \boldsymbol{p} - \iGAj \boldsymbol{q} + \iGAj d } = \rme^{ 2 c + 2 \iGAj d } = \rme^{ 2 c} \rme^{ 2 \iGAj d }.
\ee
The unitary condition $ |K |^2 =  1 $ then requires $ c = 0 $, $ d = n \pi $, $ n $ an integer.  Hence $ K = \rme^{  \boldsymbol{p} + \iGAj \boldsymbol{q}} \rme^{  \iGAj n \pi } = \pm \rme^{  \boldsymbol{p} + \iGAj \boldsymbol{q}} $. Hence, as the bilinear product $ K \cliffconj{K} $ will be unaffected, we can simply set $ n = d = 0 $.
Then, writing $ L = \rme^{ \boldsymbol{r} + \iGAj \boldsymbol{s} } $, with  $ \boldsymbol{r},\boldsymbol{s} \in \Re^3 $, we finally produce the general transformation operation
\be \label{HomLorentzGroup}
M' = \rme^{ \boldsymbol{p} + \iGAj \boldsymbol{q}  } M \rme^{ \boldsymbol{r} + \iGAj \boldsymbol{s} },
\ee
which will leave the multivector amplitude invariant.  The four three-vectors $ \boldsymbol{p}, \boldsymbol{q}, \boldsymbol{r}, \boldsymbol{s} $ illustrate that the set of transformations is a twelve dimensional manifold, thus generalizing the conventional six dimensional Lorentz group, consisting of boosts and rotations. For comparison, the conventional Lorentz transformations can be written as
\be \label{ConventionalHomLorentzGroup}
M' = \rme^{ -\boldsymbol{p} - \iGAj \boldsymbol{q}  } M \rme^{ -\boldsymbol{p} + \iGAj \boldsymbol{q} }.
\ee

As we can see, by comparing Eq.~(\ref{ConventionalHomLorentzGroup}) and Eq.~(\ref{HomLorentzGroup}), due to the expanded spacetime arena we are able to double the size of the conventional Lorentz group.  This obviously has a variety of significant consequences, such as new effects on space and time, which we now wish to explore further.  This also provides a generalised view of the key property of Lorentz invariance.

Now that we have identified $ M \cliffconj{M} $ as invariant under the set of transformations in Eq.~(\ref{HomLorentzGroup}), it is now relevant to look for other invariant expressions following from this.

\subsection{An invariant multivector dot product}

Since $ M \cliffconj{M} $ is invariant, then $ (A + B)(\overline{A + B}) $ must also be invariant, where $ A,B \in C\ell(\Re^3)$. We have
\be
(A + B)(\overline{A + B}) = A \cliffconj{A} + B \cliffconj{B} + A \cliffconj{B} + B \cliffconj{A}  .
\ee
Hence, as $ A \cliffconj{A}, B \cliffconj{B} $ are known to be invariant, then we can define a multivector dot product with the final two terms
\be \label{MultivectorDotProduct}
A \cdot \cliffconj{B} = \frac{1}{2} \left ( A \cliffconj{B} + B \cliffconj{A} \right ) = B \cdot \cliffconj{A} .
\ee
This is also an invariant,  being in the form of a complex-like number. 
The invariant dot product thus provides a mechanism to combine two distinct multivectors, as in the electromagnetic Lagrangian $ A \cdot \cliffconj{J} $, for example.

\subsection{Multivectors formed by a product defined as fields}

Now, multivectors formed from a product of two multivectors $ \cliffconj{A} B $ transform as
\be \label{multivectorProductCliffConj}
 \cliffconj{A}' B'  = \overline{K A L} K B L  = \cliffconj{L} \cliffconj{A} \cliffconj{ K} K B L  =   \cliffconj{L }  \cliffconj{A} B L  .
\ee
Hence multivectors formed as a product $ F = \cliffconj{A} B $ form a distinct class of multivectors with a distinct transformation law 
\be
F' = \cliffconj{L } F L .
\ee
We will refer to such quantities as ``fields'', as we  find this transformation applies to the electromagnetic field, for example.  
We find that the product of two fields $ F_1' F_2'= \cliffconj{L} F_1 L \cliffconj{L} F_2 L  = \cliffconj{L} (F_1 F_2) L $, also transforms as a field.

Hence, physical quantities that are produced as a product  $ F = \cliffconj{A} B $, are found to obey the conventional Lorentz transformations of Eq.~(\ref{ConventionalHomLorentzGroup}).  This being the case for the conventional electromagnetic field, for example, and hence we do not predict any novel behaviour for this case. It also appears to be confirmation of our generalised approach that we have nevertheless produced the correct transformation for the electromagnetic field. However, we do expect novel behaviour from the spacetime background, as well as the electromagnetic potential, for example, as they will be subject to the more general transformations in Eq.~(\ref{HomLorentzGroup}).  This leads to the expectation of magnetic monopoles, as we show in Section 3.7.

\subsection{Reproducing the conventional Lorentz group}

Now, if we represent spacetime by the multivector in Eq.~(\ref{multivector3DSimple}), then simple rotations of this space are described by the special case of Eq.~(\ref{HomLorentzGroup}) 
\be
M' = \rme^{ - \iGAj \boldsymbol{w}/2 } M \rme^{ \iGAj \boldsymbol{w}/2 } ,
\ee
which will produce a rotation of $ \theta = || \boldsymbol{w} || $ radians about the axis $ \boldsymbol{w} $.  Where we write the Pythagorean length of a vector $ \boldsymbol{w} $, by the unbolded symbol $ w =  \sqrt{ {\boldsymbol{w}^2}} = \sqrt{w_1^2 +w_2^2 + w_3^2}  $.

A further special case of Eq.~(\ref{HomLorentzGroup}) is found by selecting the vector exponent, which will correspond to conventional Lorentz boosts.  That is
\be \label{boostsMultivector}
M' = \rme^{ -\phi \hat{\boldsymbol{v}}/2 } M \rme^{ -\phi \hat{\boldsymbol{v}}/2 },
\ee
where $ \phi $ is defined through $ \tanh \phi = v $ where $ v = ||\boldsymbol{v} ||  $. We can rearrange $ \tanh \phi = v $ to give  $ \cosh \phi = \gamma $ and $ \sinh \phi = \gamma v  $. Hence $  \rme^{ -\phi \hat{\boldsymbol{v}} } =  \cosh \phi - \hat{\boldsymbol{v}} \sinh \phi = \gamma \left ( 1 - \boldsymbol{v} \right ) $, where $ \gamma = 1/\sqrt{1 - \boldsymbol{v}^2} $. In special relativity,  the vector $ \boldsymbol{v} $ is identified with the relative velocity vector between frames whereas for rotations it is identified with the rotation axis.  

If we now consider the effect of a Lorentz boost on the generalized eight-dimensional spacetime coordinate $ M = a +  \boldsymbol{x}_{\parallel}  + \boldsymbol{x}_{\perp} + \iGAj \boldsymbol{n}_{\parallel}  + \iGAj \boldsymbol{n}_{\perp} + \iGAj b $, where we split the spatial coordinate into components perpendicular and parallel to the boost direction $ \hat{\boldsymbol{v}} $, then we  find from Eq.~(\ref{boostsMultivector}) that
\bea \label{XBoostCoordinateMultivector}
M' & = & a \rme^{ -\hat{\boldsymbol{v}} \phi } + \boldsymbol{x}_{\parallel} \rme^{ -\hat{\boldsymbol{v}} \phi}  + \boldsymbol{x}_{\perp} \\ \nonumber
& & + \iGAj \boldsymbol{n}_{\parallel} \rme^{ -\hat{\boldsymbol{v}} \phi}  + \iGAj \boldsymbol{n}_{\perp}  + b \iGAj \rme^{ -\hat{\boldsymbol{v}} \phi } \\ \nonumber
& = &  \gamma \left ( a  - v  x_{\parallel} \right ) + \gamma \left ( \boldsymbol{x}_{\parallel} - \boldsymbol{v} a \right ) + \boldsymbol{x}_{\perp} \\ \nonumber
& & + \iGAj \gamma \left ( \boldsymbol{n}_{\parallel} - \boldsymbol{v} b \right ) + \iGAj \boldsymbol{n}_{\perp} + \iGAj \gamma \left ( b  -  v  n_{\parallel} \right ) ,
\eea
which now shows the transformation of the eight-dimensional multivector subject to the conventional Lorentz boost operation.
We can see that the plane $ \iGAj \boldsymbol{n}_{\parallel} $ orthogonal to the boost direction $ \boldsymbol{v} $ is expanded by the $\gamma $ factor to $  \iGAj \gamma \boldsymbol{n}_{\parallel} $.  This implies that the bivectors do not refer to quantities such as angular momentum of extended bodies---as the parallel components are in fact unchanged by such boosts---but must refer to axial vector-type quantities such as spin or the magnetic field. 
In fact, the bivector and trivector components $ \iGAj \boldsymbol{n} + \iGAj b = \iGAj (b + \boldsymbol{n} ) $ are transformed the same as a four-vector and indeed have the same transformational properties as the spin four-vector. 

We can also see that the conventional Lorentz boost transformation splits the multivector space into two disjoint four dimensional subspaces $  \Re \oplus \Re^3  $ and $  \bigwedge^2 \Re^3 \oplus \bigwedge^3 \Re^3  $ represented by $ a +  \boldsymbol{x} $ and $ \iGAj \boldsymbol{n}  + \iGAj b $ respectively. The first four-vector $  a +  \boldsymbol{x} $ can be identified as conventional spacetime if we identify the scalar $ a $ with the time $ t $ and the second four-vector $ \iGAj \left (\boldsymbol{n}  +  b \right ) $ as four-spin.  Thus Eq.~(\ref{multivector3DSimple})  describes a unified formulation of spacetime that incorporates spin.
 The fact that the conventional boost operation, shown in Eq.~(\ref{XBoostCoordinateMultivector}),  effectively splits the multivector into two independent four-vector spaces also illustrates why the four-vector notation is generally sufficient. However, if we wish to include the more general transformations, then we will require the eight dimensional multivector.  
Note that Eq.~(\ref{HomLorentzGroup}) combines the boost and rotation operations into a single operator and represents the relativistic effect called Thomas rotation~\cite{chappell2011revisiting}.

Hence, we can write a generalised spacetime event $ X $, in differential form, as
\be \label{multivectorMetric}
d X = d t + d \boldsymbol{x} + \iGAj d \boldsymbol{n} + \iGAj d b ,
\ee
where the special case $ d X = d t + d \boldsymbol{x} $ is isomorphic to the conventional Minkowski four vector $ d X = [ d t , d \boldsymbol{x} ] $.
Referring to Eq.~(\ref{AmplitudeSquared}) this therefore has  amplitude  
\be \label{SpacetimeAmplitudeSquared}
| d X |^2 = d t^2 - d \boldsymbol{x}^2 + d \boldsymbol{n}^2 - d b^2 + 2 \iGAj \left (d t d b - d \boldsymbol{x} \cdot d \boldsymbol{n}  \right ) .
\ee
In general, the invariant interval is therefore a complex-like number, as it now contains an additional imaginary term.  
Hence, based on the requirement for the most general invariant quantity in $ C\ell(\Re^3) $ we see that the Minkowski line element $ dt^2 - d \boldsymbol{x}^2 $ has naturally arisen, as well as a generalisation. The origin of the Minkowski metric therefore appears to be the operation of Clifford conjugation acting on physical space, as modelled by $ C\ell(\Re^3) $.  Clifford conjugation, shown in Eq.~(\ref{CliffordConjugation}), which reverses the linear motion and spin directions, thus appears equivalent to a time reversal on the space.  That is, velocities and spins are reversed whereas scalars and helicity are unchanged.  Hence, spacetime combined with its time reversed copy $ d X d \cliffconj{X} $, is what appears to create the Minkowski spacetime structure.

We now wish to explore some of the immediate consequences of the generalised spacetime metric in Eq.~(\ref{SpacetimeAmplitudeSquared}).

\section{Results}

\subsection{Proper time is two dimensional}

  For the general metric in Eq.~(\ref{SpacetimeAmplitudeSquared}), in order to change to a comoving frame (in order to measure its proper time), we not only need to co-move but also co-rotate with the frame.  This nevertheless leaves two quantities, the scalar time and the helicity. The helicity describes a form of twisting of spacetime at each point (like the wringing of a towel).  Hence, time becomes intrinsically two-dimensional, with a distinct geometrical nature, combining scalar and pseudoscalar aspects.  
  That is we could write the proper time as  $ d \tau =  dt + \iGAj d b $.
Recently, this idea of more than one time dimensions has been given validity by experiment, with two-dimensional time being created within a quantum computer~\cite{Dumitrescu2022}.  Also previous theoretical work on two time dimensions has shown it to be a physically meaningful hypothesis~\cite{bars2001survey}.

Now, as we find the concepts of time and space emerging naturally from $ C\ell(\Re^3) $, we can seek to gain a greater insight into their underlying properties.
The invariant distance $ d t^2 - d \boldsymbol{x}^2 $ effectively means that each observer will see a spherically expanding light shell of radius $ |\boldsymbol{x}| = t $. Hence, due to the spherical symmetry for all observers, a single number $ t $, can describe its radius.  In comparison, we require three dimensions of space in $ C\ell(\Re^3) $.  
  Hence, this provides an explanation for why time appears one dimensional whereas space is three dimensional and why it is possible to freely move in the space dimensions but not in the time dimension, which naturally describes a spherical outward expansion from a point.    Also, we notice in Eq.~(\ref{SpacetimeAmplitudeSquared}), that an additional non-squared time factor $ dt db $ arises in the imaginary component, which breaks the normal symmetry in the time direction, thus also giving an arrow to time, which is lacking in the Minkowski metric.

\subsubsection{Reducing the proper time along a worldline}

The true time elapsed for a travelling observer, is typically given by the proper time $ d \tau^2 = dt^2 - d \boldsymbol{x}^2 $.  
From Eq.~(\ref{SpacetimeAmplitudeSquared}), we can see that by setting the spin term $ d \boldsymbol{n} =0 $, we can reduce this proper time by increasing the helicity, giving
\be \label{SpacetimeAmplitudeSquaredMinimiseProperTime}
d \tau^2 = d t^2 - d \boldsymbol{x}^2 - d b^2 + 2 \iGAj d t d b  .
\ee
Hence, the scalar time elapsed is now $ d t^2 - d \boldsymbol{x}^2 - d b^2 $, indicating that it can be reduced by increasing the helicity $ b $.
We can see that the two time dimensions interact as $ d t^2 - d b^2 $, allowing a reduction in the proper time, using the second time dimension, $ b $.  We also note that the proper time distance has an imaginary component, which needs to be interpreted.

This shows that the twin paradox can be generalised, and that the elapsed time of the travelling twin can perhaps be reduced beyond what is derived in SR.   This is not the only interpretation of the effect since conditions other than $ d \boldsymbol{n} =0 $ can set the initial conditions.

\subsection{Velocity multivector}

The magnitude of the invariant interval in Eq.~(\ref{SpacetimeAmplitudeSquared}), is commonly defined equal to $ d \tau^2 $, which defines the proper time.  
Dividing through by this invariant, from Eq.~(\ref{multivectorMetric}), we produce the velocity multivector  
\bea \label{velocityMultivector}
V & = & \frac{d X}{d \tau} = \frac{d t}{d \tau } + \frac{d \boldsymbol{x}}{ dt }\frac{ dt }{d \tau} + \iGAj \frac{d \boldsymbol{n}}{d t} \frac{d t}{d \tau } + \iGAj \frac{d b}{d t}\frac{d t}{d \tau} \\ \nonumber
& = & \gamma \left ( 1 + \boldsymbol{v} + \iGAj  \boldsymbol{w} + \iGAj h \right ) , \nonumber
\eea
where $\boldsymbol{v} = \frac{d \boldsymbol{x}}{d t} $, $\boldsymbol{w} = \frac{d \boldsymbol{n}}{d t} $ and $h= \frac{d b}{d t}$.
As we defined $ | d X |^2 = d \tau^2 $ then we have $ |V|^2 = \frac{| d X |^2}{d \tau^2} = 1 $, a dimensionless number.  
Therefore
\be
|V|^2  = \gamma^2 \left ( 1 - \boldsymbol{v}^2 +  \boldsymbol{w}^2 - h^2  + 2 \iGAj \left (h -  \boldsymbol{v} \cdot  \boldsymbol{w}  \right )\right ) = 1
\ee
and hence
\be \label{fullGammExpression}
\gamma =  \frac{d t}{d \tau} = \frac{1}{\sqrt{\left ( 1 - \boldsymbol{v}^2 +  \boldsymbol{w}^2 - h^2  + 2 \iGAj \left (h -  \boldsymbol{v} \cdot  \boldsymbol{w}  \right )\right )}} .
\ee
This then generalises the time dilation factor to account for spin $ \boldsymbol{w} $ and helicity $ h $.

Now, the conventional special relativistic $ \gamma = \frac{1}{\sqrt{1-v^2}} $, becomes imaginary if $ v > 1 $, and so is generally regarded as unphysical.
However, for the general velocity multivector in Eq.~(\ref{velocityMultivector}), there is no difficulty with an imaginary gamma factor, as it simply implies that the linear motion $\boldsymbol{v}$ is converted into spinning motion $ \iGAj \boldsymbol{w}$ (and vice versa). Hence, this expanded eight-dimensional framework is able to describe superluminal particles~\cite{PhysRev.159.1089,recami1978tachyons}, in a natural manner.

Also, as $ |V|^2 = 1 $, then it is naturally expressed in exponential form
\be
V = \rme^{\phi \frac{\boldsymbol{v} + \iGAj  \boldsymbol{w}}{\sqrt{\left (\boldsymbol{v} + \iGAj  \boldsymbol{w} \right )^2} } } = \cosh \phi +  \frac{\boldsymbol{v} + \iGAj \boldsymbol{w} }{\sqrt{(\boldsymbol{v} + \iGAj \boldsymbol{w})^2 } } \sinh \phi .
\ee
The exponential form shows that a change in eight-velocity involves a hyperbolic rotation of the  multivector in eight dimensions.  Hence, the eight-velocity takes a dual role, of representing velocity as well as acting as an operator to change frames, according to Eq.~(\ref{HomLorentzGroup}).

Now, as $ |V|^2 = 1 $ is constant, then differentiating, we find $ \frac{d }{d \tau} |V|^2 = \frac{d V}{d \tau} \cliffconj{V}  + V \frac{d \cliffconj{V}}{d \tau} = 0 $, using the product rule of differentiation.  So, defining $ A = \frac{d V}{d \tau} $ for an acceleration multivector, we thus produce an orthogonality condition for the velocity and acceleration multivectors as $ A \cliffconj{V} + V \cliffconj{A} = 0 $ or $ A \cdot \cliffconj{V} = 0 $, a generalisation of the conventional four-vector orthogonality result for eight-vectors.

We note that we are not addressing accelerating frames in this paper, and this possible extension does need to be treated carefully~\cite{coleman2017bell}.  Nevertheless, we believe this new derivation in the context of inertial frames, which attempts to reveal the deeper origins of Minkowski spacetime is of sufficient importance to describe on its own merit.  The extension to accelerating frames is an open question for future work.

\subsection{The gradient}

For a multivector $ M(t,x_1,x_2,x_3,n_1,n_2,n_3,b) $, varying over its eight dimensions, we define the gradient:
\begin{defn}[Eight gradient]
We define the gradient operator
\be
\partial = \frac{\partial}{\partial t}  +\nabla + \iGAj e_1 \frac{\partial}{\partial n_1} + \iGAj e_2 \frac{\partial}{\partial n_2}+ \iGAj e_3  \frac{\partial}{\partial n_3}+ \iGAj \frac{\partial}{\partial b} ,
\ee
where $ \nabla =  e_1 \frac{\partial}{\partial x_1} + e_2 \frac{\partial}{\partial x_2}+  e_3  \frac{\partial}{\partial x_3} $, is the regular three-gradient.
\end{defn}

We thus have the special case of the four-gradient varying just over the time and space dimensions, with
\be \label{FourGradient}
\partial = \frac{\partial}{\partial t} + e_1 \frac{\partial}{\partial x} + e_2 \frac{\partial}{\partial y} + e_3 \frac{\partial}{\partial z} 
\ee
giving 
\be  \label{WaveOperator}
\partial \cliffconj{\partial} = \frac{\partial^2}{\partial t^2} - \frac{\partial^2}{\partial x^2} - \frac{\partial^2}{\partial y^2} - \frac{\partial^2}{\partial z^2} ,
\ee
the standard wave operator.

\subsection{Elementary equations}

Now, the product of a multivector with a field  $ X F $ will transform the same as a general multivector.  That is $ X' F' = K X L \cliffconj{L} F L = K (X F) L $. 
Hence, we can write an invariant equation $ X F = Y $, where $ X, Y $ transform as multivectors, defined in Eq.~(\ref{HomLorentzGroup}), and $ F = \cliffconj{B} A  $ transforms as a field.  

Now, given a quantity derived from the gradient of a multivector potential $ A $ as $ F = \cliffconj{\partial} A $, this then being a product of two multivectors, matches our definition of a field. 
Selecting $ X = \partial $, $ F = \cliffconj{\partial} A  $ and $ Y = J $, we produce the general form of Maxwell's equations~\cite{Chappell2014IEEE,Baylis2001}, $ \partial F = J $, where $ J $ represents the sources. We find that this is a generalisation of Maxwell's equations, as fully populating the source term over the multivector $ J  $, will include magnetic monopole sources, as shown in~\ref{MaxwellAppendix}.

The field transformation $  F' = \cliffconj{L} F L $ or
\be
(\boldsymbol{E} + \iGAj \boldsymbol{B})' = \rme^{ -\boldsymbol{r} - \iGAj \boldsymbol{s} } (\boldsymbol{E} + \iGAj \boldsymbol{B}) \rme^{ \boldsymbol{r} + \iGAj \boldsymbol{s} } ,
\ee
turns out to be the standard transformation for the electromagnetic field~\cite{Chappell2014IEEE}.
Hence, our generalized transformation for spacetime shown in Eq.~(\ref{HomLorentzGroup}) leaves Maxwell's equations invariant as well as retaining the conventional field transformation.  Indeed, the Lorentz transformations were originally developed as the transformations that leave Maxwell's equations invariant~\cite{einstein1905electrodynamics}, and we therefore have found a more general class of such transformations, as shown in Eq.~(\ref{HomLorentzGroup}).

As an extension, another elementary equation we could write is  
$ \partial F  =  Y F^*  $, where $ Y, F $ are eight vectors, which is equivalent to the Dirac equation.  The eight-dimensional multivector $ F $, naturally corresponding with the eight-dimensional Dirac spinor.  This description also makes it explicit that the source free Maxwell equation $ \partial F  = 0 $ is isomorphic to the massless Dirac equation.

\subsection{Cartesian rotations in 4D}

If we consider the transformation
\be \label{rotationInTime}
M' = \rme^{ \iGAj \boldsymbol{v}/2 } M \rme^{ \iGAj \boldsymbol{w}/2 },
\ee
where we have used two distinct rotation axes $  \boldsymbol{v} $ and $  \boldsymbol{w} $.
It can be shown, for the multivector given in Eq.~(\ref{multivector3DSimple}), that this operation acts separately on  two four-dimensional subspaces $ t + \iGAj \boldsymbol{n} $ and $ \boldsymbol{x} + \iGAj b $, with each of the two rotations being isomorphic to a rotation in a four-dimensional Cartesian space. This transformation, even though it preserves the invariant distance, is not able to be included in the standard Lorentz group as it lies outside the conventional four-vector representation.  

We can see that this transformation can move space and time quantities into or out of the spin terms in the multivector.  This can be further explored for modeling  physical processes that involve various space and time dilations.

\subsection{Lightlike particles}

Lightlike particles satisfy the condition $  dt^2 - d \boldsymbol{x}^2 = 0 $.  If we now also enforce a null condition $ |d X|^2 = 0 $, using the generalized metric in Eq.~(\ref{SpacetimeAmplitudeSquared}), and consider a light speed particle with $ d t^2 - d \boldsymbol{x}^2 = 0 $, then from Eq.~(\ref{SpacetimeAmplitudeSquared}), we require $ d b^2= d \boldsymbol{n}^2 $ or $ d b= \pm || d \boldsymbol{n} || $ and $ d b d t - d \boldsymbol{x} \cdot d \boldsymbol{n}=0  $. By combining these two results we produce  the condition $ d \boldsymbol{x}\cdot d \boldsymbol{n}= \pm || d \boldsymbol{n} || t $. Dividing through by $ || d \boldsymbol{n} || $ gives $
d \boldsymbol{x}\cdot \hat{\boldsymbol{n}}= \pm d t $,
where $ \hat{\boldsymbol{n}} $ is the unit vector in the direction of $ \boldsymbol{n} $ and so is a relation describing Einstein's light cone. Now, $ d \boldsymbol{x} $ has its minimum value of  $ d t $ when it is parallel to $ \hat{\boldsymbol{n}} $ and so this relation enforces a space-like condition.  We then find the general condition for null lightlike particles to be
\be
\boldsymbol{v}\cdot \hat{\boldsymbol{n}}= \pm c  ,
\ee
where for clarity we introduce the speed of light. Hence, due to the nature of the dot product, we can see that it is only satisfied by a velocity $ || \boldsymbol{v} || = c $,  parallel to the spin axis $ \hat{\boldsymbol{n}} $. That is, based on the eight-dimensional structure of $ C\ell(\Re^3) $ alone, we find that a null particle, if traveling at the speed of light $ c$, is required to have its spin axis parallel to its direction of motion,  exactly as observed for electromagnetic radiation.  This result, though, being derived here using purely algebraic arguments from $ C\ell(\Re^3) $.

\subsubsection{General null particles}

More generally, if we do not explicitly enforce a light velocity for null particles, but only the requirement that $ |d X|^2 = 0 $, then we have the equation
\be \label{generalNullRelation}
0 =   1 - \boldsymbol{v}^2 + \boldsymbol{w}^2  - h^2 + 2 \iGAj \left(  h -  \boldsymbol{v} \cdot \boldsymbol{w} \right ) .
\ee
We require $ h =  \boldsymbol{v} \cdot \boldsymbol{w} $ in order to zero the imaginary component, and so we have 
\be \label{generalSpeedLight}
0 =   1 - \boldsymbol{v}^2 + \boldsymbol{w}^2  - \left ( \boldsymbol{v} \cdot \boldsymbol{w} \right )^2 =  1 - \boldsymbol{v}^2 + \boldsymbol{w}^2  - \cos^2 \theta \, \boldsymbol{v}^2 \boldsymbol{w}^2 ,
\ee
which gives
\be \label{superLuminal}
||\boldsymbol{v} ||= \pm c \sqrt{\frac{ 1  + \boldsymbol{w}^2}{1+\cos^2 \theta \,  \boldsymbol{w}^2 }}.
\ee
We recover the previous result, that a particle must travel at the speed of light if its spin axis is parallel to its velocity, that is with $ \theta = 0 $.
However, if we can generate radiation, with the spin axis orthogonal to its propagation direction, we would achieve a velocity of $ ||\boldsymbol{v} ||= \sqrt{ 1  + \boldsymbol{w}^2} $, which therefore indicates the possibility of superluminal light propagation~\cite{dragan2022relativity}.  This orthogonal  property has now been observed in near field electromagnetic radiation~\cite{bliokh2016transverse,PhysRev.159.1089,recami1978tachyons}, but we would also predict that this radiation will be superluminal, based on Eq.~(\ref{superLuminal}).
As this radiation is in the evanescent near field and non-propagating it needs specific experimental testing, some of which indeed appear to show faster than light transmission~\cite{kholmetskii2007measurement}.
This formalism could therefore have application in areas where faster than light propagation appears to occur, such as in quantum tunneling, for example.

Superluminal propagation is a debated topic, but it is generally recognised in phase waves and has also been theoretically predicted and observed in quantum tunneling~\cite{dumont2020relativistic}. There is  ongoing scientific debate regarding the nature of this superluminal behavior and on whether it extends to superluminal signaling and the affect on causality~\cite{withayachumnankul2010systemized,dumont2020relativistic}.
We are careful to describe here in this paper a specific form of wave generation that we predict would be superluminal.  It is possible, that this is not generated in Nature, and so there is no conflict with current experiments.

\subsection{Magnetic monopoles}

For an electromagnetic four-potential $ \phi + \boldsymbol{A} $, subject to the generalised transformations in Eq.~(\ref{HomLorentzGroup}), will create terms across the whole multivector, as $ \phi' + \boldsymbol{A}' + \iGAj \boldsymbol{M}' + \iGAj \psi' $, where $ \iGAj \boldsymbol{M}' $ will represent a magnetic monopole current potential and $ \iGAj \psi' $ is a magnetic monopole source.  Hence, the generalised transformations appear to indicate the possible existence of magnetic monopoles.

While free magnetic monopoles have not yet been experimentally detected, their emergent properties have been observed in spin ice materials~\cite{grigera2010dirac}.

\section{The action}

The Lorentz invariant distance provides a suitable action integral
\be \label{ActionMetricDistance}
S =  \int | d X | ,
\ee
where the distance $  | d X | $ is given by the amplitude of the spacetime multivector, given by Eq.~(\ref{AmplitudeSquared}). That is, we are following the standard procedure of extremizing the proper time in order to find the geodesics. Now, as shown previously, with the assumption of a proper time in a rest frame we have $ | d X | = d \tau $ and so we have the spacetime distance
\be
|dX|^2 = \left ( \dot{t}^2 - \dot{\boldsymbol{x}}^2 + \dot{\boldsymbol{n}}^2 - \dot{b}^2  \right ) d \tau^2,
\ee
where we define $ \dot{t} = \frac{d t}{d \tau} $, $ \dot{\boldsymbol{x}} =  \frac{d \boldsymbol{x}}{d \tau} $, $ \dot{\boldsymbol{n}} =  \frac{d \boldsymbol{n}}{d \tau} $ and $ \dot{b} = \frac{d b}{d \tau} $.  We can then write the action as $ S = \int \frac{| d X | }{d \tau} d \tau $ that implies a Lagrangian
\be \label{LagrangianVelocity}
\mathcal{L} = \frac{| d X | }{d \tau} = |V|  = \sqrt{ \dot{t}^2 - \dot{\boldsymbol{x}}^2 + \dot{\boldsymbol{n}}^2 - \dot{b}^2 } = 1,
\ee
where we now extremize  $ S = \int \mathcal{L} d \tau $.

As we have no explicit coordinate dependence,  $ \frac{ \partial \mathcal{L}}{\partial \dot{t} } $, $  \frac{ \partial \mathcal{L}}{\partial \dot{\boldsymbol{x}} } $, $  \frac{ \partial \mathcal{L}}{\partial \dot{\boldsymbol{n}} } $ and $ \frac{ \partial \mathcal{L}}{\partial \dot{b} } $ are constants of the motion.
Using the Euler-Lagrange equation~\cite{goldstein2002} for $ t $ 
\be
\frac{d}{d \tau} \frac{ \partial \mathcal{L}}{\partial \dot{t} } = \frac{ \partial \mathcal{L}}{\partial t }  = 0
\ee
thus giving the conserved quantity
\be  \label{energyInvariantLagrangian}
\frac{ \partial \mathcal{L}}{\partial \dot{t} } = \mathcal{L}^{-1} \dot{t} = E  .
\ee
We have written the conserved quantity $ E $ as we expect it to relate to energy by Noether's theorem.  Indeed, because $  \dot{t} = d t/d \tau = \gamma $ and $ \mathcal{L}^{-1} = 1 $, we find equating real components that $ E = \gamma  $.  
The second conserved quantity will be $ \boldsymbol{p} = \gamma  \boldsymbol{v} $.  

The bivector component will produce the conservation of relativistic angular momentum $ \boldsymbol{s} = \gamma  \boldsymbol{w} $ as expected and 
the fourth conserved quantity will be
\be
\frac{ \partial \mathcal{L}}{\partial  \dot{b}  } = \mathcal{L}^{-1} \dot{b} = H,
\ee
that returns the helicity $ H =  \dot{b} = \gamma   h  $.  

Thus the multivector invariant interval in Eq.~(\ref{ActionMetricDistance}), encodes the four fundamental conservation laws for inertial particles~\cite{lehmkuhl2014einstein}. 

As well as unifying four conservation laws, it also indicates that spin and helicity (as well as energy and momentum) are intrinsic to spacetime.  The conservation of momentum obviously being a restatement of Newton's first law.

\subsection{Electromagnetic Lagrangian}

We noted that the electromagnetic field is naturally defined within this formalism as $ F = \cliffconj{\partial} A $, as opposed to the conventional $ F = \partial \wedge A $. This form was first proposed by Fermi as an alternative electromagnetic Lagrangian~\cite{Oosten2000}.
The advantage of using the definition $ F = \cliffconj{\partial} A $,  is that
\bea
F & = & \left ( \frac{\partial}{\partial t}  - \nabla \right ) \left ( \phi - \boldsymbol{A} \right ) \\ \nonumber
& = & \frac{\partial \phi }{\partial t} + \nabla \cdot \boldsymbol{A} - \nabla \phi - \frac{\partial \boldsymbol{A}}{\partial t} + \nabla \wedge \boldsymbol{A} \\ \nonumber
& = & \ell + \boldsymbol{E} + \iGAj \boldsymbol{B} , \nonumber
\eea
where $ \boldsymbol{E} = - \nabla \phi - \frac{\partial \boldsymbol{A}}{\partial t} $,  $ \iGAj \boldsymbol{B} = \nabla \wedge \boldsymbol{A} = \iGAj \nabla \times \boldsymbol{A} $ and $ \ell = \frac{\partial \phi }{\partial t} + \nabla \cdot \boldsymbol{A} $. This thus automatically gives the correct field definitions as well as the Lorenz gauge $ \ell $.  In order to recover the standard electromagnetic field $ F = \boldsymbol{E} + \iGAj \boldsymbol{B} $ we need to adopt the Lorenz gauge with $ \ell = 0 $. The Lorenz gauge produces a Lorentz invariant form of electromagnetism, enforcing causality and charge conservation, which is generally assumed to be a requirement of a physical theory.  

Now, with $ \ell = 0 $, we find $ F \cliffconj{F} = - F^2 $. Specifically
\be
F^2 = (\boldsymbol{E} + \iGAj c \boldsymbol{B} ) (\boldsymbol{E} +\iGAj c \boldsymbol{B} ) =  \boldsymbol{E}^2 - \boldsymbol{B}^2 + 2 \iGAj \boldsymbol{E} \cdot \boldsymbol{B}.
\ee
Therefore, we can write an electromagnetic field Lagrangian
\be
\mathcal{L} =  \frac{1}{2} F^2  + A \cdot \cliffconj{J} = \frac{1}{2}   (\cliffconj{\partial} A)^2  + A \cdot \cliffconj{J}.
\ee
Varying $ \mathcal{L} $ with respect to $ A $ produces Maxwell's equations $ \partial \cliffconj{\partial} A =\partial F = J $.
This form of the electromagnetic Lagrangian has the advantage of producing a symmetric energy momentum tensor and conserved spin~\cite{Oosten2000}.

\subsection{Particle in an electromagnetic field} 

We found the Lagrangian for inertial particles in Eq.~(\ref{LagrangianVelocity}), of $ \mathcal{L} =  |V| $, being the dimensionless magnitude of the eight velocity. The simplest extension of this Lagrangian, while maintaining invariance could possibly  be $ \mathcal{L} =  |V + U | $, where the multivector $ U $ conceptually represents a `flow' in the background spacetime, perturbing particle inertial motion $ V $.  We can also add the known invariant of $ 
 V \cdot \cliffconj{A} $ to the Lagrangian.
We thus produce a generalised Lagrangian
\be \label{GravityLagrangian}
\mathcal{L} =  \frac{1}{2}| V+ U|^2  + A \cdot \cliffconj{V} .
\ee
Note that we are permitted to use either $ \mathcal{L} =  |V + U| $ or $ \mathcal{L} =  \frac{1}{2} |V + U |^2 $, because if a Lagrangian $ \mathcal{L} $ satisfies the Euler-Lagrange equations, then in general any function  $ F(\mathcal{L}) $ of the Lagrangian also satisfies the Euler-Lagrange equations.  
We can also recognize $ A \cdot \cliffconj{V}  $ as being in the form of the conventional classical electromagnetic Lagrangian.

The term $ \frac{1}{2}| V+ U|^2 $ including an additional deflection from inertial motion and its physical consequences is an open question for future work.

\section{Conclusion}

We produce Minkowski spacetime as an emergent property of physical space, when modeled by the Clifford algebra $ C\ell(\Re^3) $, shown in Eq.~(\ref{multivectorMetric}).
 The interaction of the four algebraic elements in $ C\ell(\Re^3) $ of scalars, polar vectors, axial vectors and helicity, which represent four types of physical quantities, produce the Minkowski metric.  These four quantities, also geometrically represent  points, lines, areas and volumes, thus giving a new geometric basis to Minkowski spacetime.
Also, as $ C\ell(\Re^3) $ has eight degrees of freedom, we naturally produce a generalised eight dimensional spacetime arena.  We find that the additional four degrees of freedom describe the properties of spin and helicity. 

Einstein developed special relativity from first principles, based on the relativity of uniform motion and the fixed speed of light,  which then led to the Minkowski metric.
In distinction with our approach, we find that the Minkowski spacetime structure, including the fixed speed of light, arises naturally from the principles of geometry, when described by $ C\ell(\Re^3) $.

Einstein acknowledged that SR was a theory of principle (based on two axioms) rather than a constructive theory working from first principles.  With the assumption of $ C\ell(\Re^3) $ as a basis for spacetime, we have thus provided a more constructive approach to SR.

Our generalised invariant interval of Eq.~(\ref{SpacetimeAmplitudeSquared}) gives a set of allowable transformation equations, which coincide with  the conventional Lorentz transformations as special cases. 
We also show that the   transformations in Eq.~(\ref{HomLorentzGroup}) are a more general class yet still hold Maxwell's equations invariant. This allows us to predict a form of electromagnetic radiation that is superluminal, provided its spin axis is not parallel with its propagation direction. The maximum speed being reached when its spin axis is orthogonal to the propagation direction, as shown in Eq.~(\ref{superLuminal}).  
We also found the generalised metric can modify the proper time on a worldline, leading to a possible generalisation of the twin paradox, for example.

There is a philosophical question regarding the independent reality of spacetime~\cite{brown2006minkowski}.  This paper, adds to this debate, as it ascribes the local Minkowskian behaviour of the gravitational field, as derived from the fundamental interaction of the four geometrically distinct elements---energy, momentum, spin and helicity, as described by $ C\ell(\Re^3) $. The gravitational metric is now a function of  coordinates themselves, which allows one to use coordinates in a neo-Kantian formal manner and the metric tensor to satisfy logical empiricism.


A further consequence of this derivation is that time becomes represented as the scalar component of the multivector. This interpretation thus obviates the need for an additional fourth Euclidean-type dimension to describe time, as it already exists within  $ C\ell(\Re^3) $. 
The invariant interval is in general a complex like number, and hence the proper time is also complex, and thus two-dimensional~\cite{bars2001survey,Cole1980comments,dorling1970dimensionality}.

The structure of $ C\ell(\Re^3)$ thus provides an explanation for the origin of Minkowski spacetime as well as a coherent framework for special relativistic physics. This produces a generalised invariant interval in Eq.~(\ref{SpacetimeAmplitudeSquared}), generalised Lorentz transformations and Maxwell's equations,   directly from the properties of the algebra of $ C\ell\left(\Re^3\right) $.  The generalised framework also predicts a range of new physical effects, including the possibility of super-luminal light propagation. This theory proposing that local Minkowskian spacetime is an inherent property of three-dimensional physical space, also informs the Kantian and logical empiricist debate on the nature of spacetime.

\section*{Acknowledgment}

We gratefully acknowledge helpful discussions with Alex Dinovitser and insightful comments provided by Miroslav Josipovic.

This work was support by the Australian Research Council grant DP200103795.
N. Iannella's contribution was supported by the People Programme (Marie Curie Actions) of the European Union’s Seventh Framework Programme (FP7/2007-2013) under REA grant agreement No PCOFUND-GA-2012-600181

\appendix

\section{Maxwell's equations} \label{MaxwellAppendix}

It is known, that using Clifford geometric algebra, Maxwell's four equations can be written with the single equation~\cite{Chappell2014IEEE,Baylis2001}
\be \label{MaxwellExpandedGA}
\left (\frac{\partial}{\partial t} + \nabla \right) F = \rho - \mathbf{J} ,
\ee
where the electromagnetic field is represented as $  F = \boldsymbol{E} + \iGAj \boldsymbol{B}  $ and $ \nabla = e_1 \frac{\partial}{\partial x} + e_2 \frac{\partial}{\partial y} + e_3 \frac{\partial}{\partial z} $
 and the four-current $ J = \rho - \mathbf{J} $.  
Expanding this expression and equating the scalar, vector, bivector and pseudoscalar parts, we find the four Maxwell equations.

The electromagnetic field can be derived from a potential $ A $ as $ F = \cliffconj{\partial} A $ or
\bea
F & = & \left ( \frac{\partial}{\partial t}  - \nabla \right ) \left ( \phi - \boldsymbol{A} \right ) \\ \nonumber
& = & \frac{\partial \phi }{\partial t} + \nabla \cdot \boldsymbol{A} - \nabla \phi - \frac{\partial \boldsymbol{A}}{\partial t} + \nabla \wedge \boldsymbol{A} \\ \nonumber
& = & \ell + \boldsymbol{E} + \iGAj \boldsymbol{B} , \nonumber
\eea
where $ \boldsymbol{E} = - \nabla \phi - \frac{\partial \boldsymbol{A}}{\partial t} $,  $ \iGAj \boldsymbol{B} = \nabla \wedge \boldsymbol{A} = \iGAj \nabla \times \boldsymbol{A} $ and $ \ell = \frac{\partial \phi }{\partial t} + \nabla \cdot \boldsymbol{A} $, the Lorenz gauge.

\section{The multivector products}

In Clifford geometric algebra we form the space of multivectors $ \Re \oplus \Re^3 \oplus \bigwedge^2 \Re^3 \oplus \bigwedge^3 \Re^3 $, an eight-dimensional real vector space denoted by $ C\ell(\Re^3) $.  This thus consists of the sum of a scalar, vector, bivector and trivector.
Defining vectors $ \boldsymbol{v} = v_1 e_1 + v_2 e_2 + v_3 e_3 $ and $ \boldsymbol{u} = u_1 e_1 + u_2 e_2 + u_3 e_3 $, where $ v_i, v_i \in \Re $, we find their algebraic product using the distributive law of multiplication over addition as
\bea \label{VectorProductExpand}
\boldsymbol{u} \boldsymbol{v} & = & (e_1 u_1 + e_2 u_2 + e_3 u_3 ) ( e_1 v_1 + e_2 v_2 + e_3 v_3 ) \\ \nonumber
& = & u_1 v_1 + u_2 v_2 + u_3 v_3  + (u_2 v_3 - v_2 u_3 ) e_2 e_3 \\ \nonumber
& & + (u_1 v_3 - u_3 v_1 ) e_1 e_3 + (u_1 v_2 - v_1 u_2 ) e_1 e_2 \\ \nonumber
& = & \boldsymbol{u} \cdot \boldsymbol{v}  + \boldsymbol{u} \wedge \boldsymbol{v}, \\ \nonumber
& = & \boldsymbol{u} \cdot \boldsymbol{v}  + \iGAj \boldsymbol{u} \times \boldsymbol{v}, \nonumber
\eea
which produces a sum of symmetric and antisymmetric products. We find the unifying result that the algebraic product $ \boldsymbol{u} \boldsymbol{v} $ produces a complex-like number combining the dot and cross products $ \boldsymbol{u} \cdot \boldsymbol{v}  + \iGAj \boldsymbol{u} \times \boldsymbol{v} $.

\newpage

\bibliographystyle{iopart-num}

\bibliography{GeneralisedMinkowskiSpacetime}

\providecommand{\newblock}{}
\begin{thebibliography}{10}
\expandafter\ifx\csname url\endcsname\relax
  \def\url#1{{\tt #1}}\fi
\expandafter\ifx\csname urlprefix\endcsname\relax\def\urlprefix{URL }\fi
\providecommand{\eprint}[2][]{\url{#2}}

\bibitem{Einstein1905}
Einstein A 1905 {\em Annalen der Physik\/} {\bf 322} 891--921

\bibitem{Verlinde2017}
Verlinde E~P 2017 {\em SciPost Phys.\/} {\bf 2} 016
  \urlprefix\url{https://scipost.org/10.21468/SciPostPhys.2.3.016}

\bibitem{coxeter1973regular}
Coxeter H~S~M 1973 {\em Regular Polytopes\/} (Dover Pubns)

\bibitem{Oziewicz2012}
Oziewicz Z and Page W~S 2012 {\em Proceedings of the Natural Philosophy
  Alliance\/} {\bf 9} 406--409

\bibitem{hoyle2001submillimeter}
Hoyle C~D, Schmidt U, Heckel B~R, Adelberger E~G, Gundlach J~H, Kapner D~J and
  Swanson H~E 2001 {\em Phys. Rev. Lett.\/} {\bf 86}(8) 1418--1421

\bibitem{pavsic2006}
PAV\v{S}I\v{C} M 2006 {\em International Journal of Modern Physics A\/} {\bf
  21} 5905--5956

\bibitem{Hestenes:1966}
Hestenes D 1966 {\em Spacetime Algebra\/} (New York: Gordon and Breach)

\bibitem{Baylis1996}
Baylis W~E 1996 {\em Clifford (Geometric) Algebras With Applications in
  Physics, Mathematics, and Engineering\/} (Birkh{\"a}user, Boston)

\bibitem{Chappell2015}
Chappell J~M, Iqbal A, Gunn L~J and Abbott D 2015 {\em PLOSONE\/} {\bf
  10}(e0116943)

\bibitem{Chappell2014IEEE}
Chappell J~M, Drake S~P, Seidel C~L, Gunn L~J, Iqbal A, Allison A and Abbott D
  2014 {\em Proceedings of the IEEE\/} {\bf 102} 1340--1363

\bibitem{hamilton1853lectures}
Hamilton W 1853 {\em Lectures on Quaternions\/} (Royal Irish Academy)

\bibitem{Silberstein1912}
Silberstein L 1912 {\em The London, Edinburgh, and Dublin Philosophical
  Magazine and Journal of Science\/} {\bf 23} 709--809

\bibitem{dray2015geometry}
Dray T and Manogue C~A 2015 {\em The Geometry of the Octonions\/} (World
  Scientific)

\bibitem{gogberashvili2005octonionic}
Gogberashvili M 2005 {\em Advances in Applied Clifford Algebras\/} {\bf 15}
  55--66

\bibitem{gogberashvili2016octonionic}
Gogberashvili M 2016 {\em International Journal of Geometric Methods in Modern
  Physics\/} {\bf 13} 1650092

\bibitem{gogberashvili2015geometrical}
Gogberashvili M and Sakhelashvili O 2015 {\em Advances in Mathematical
  Physics\/} {\bf 2015} 1--14

\bibitem{gogberashvili2014split}
Gogberashvili M 2014 {\em The European Physical Journal C\/} {\bf 74} 1--9

\bibitem{Hestenes2003}
Hestenes D 2003 {\em American Journal of Physics\/} {\bf 71} 691--714

\bibitem{chappell2011revisiting}
Chappell J~M, Iannella N, Iqbal A and Abbott D 2012 {\em PLOSONE\/} {\bf 7}
  e51756

\bibitem{Dumitrescu2022}
Dumitrescu P, Bohnet J, Gaebler J, Hankin A, Hayes D, Kumar A, Neyenhuis B,
  Vasseur R and Potter A 2022 {\em Nature\/} {\bf 607(7919)} 463--467

\bibitem{bars2001survey}
Bars I 2001 {\em Classical and Quantum Gravity\/} {\bf 18} 3113

\bibitem{PhysRev.159.1089}
Feinberg G 1967 {\em Phys. Rev.\/} {\bf 159}(5) 1089--1105

\bibitem{recami1978tachyons}
Morita K 1979 {\em Particle Theory Research\/} {\bf 59} 204--206

\bibitem{coleman2017bell}
Coleman B 2017 {\em Results in Physics\/} {\bf 7} 2575--2581

\bibitem{Baylis2001}
Baylis W~E 2001 {\em Electrodynamics: A Modern Geometric Approach\/}
  (Birkh{\"a}user, Boston)

\bibitem{einstein1905electrodynamics}
Einstein A {\em et~al.\/} 1905 {\em Annalen der Physik\/} {\bf 17} 891--921

\bibitem{dragan2022relativity}
Dragan A, D{\c{e}}bski K, Charzy{\'n}ski S, Turzy{\'n}ski K and Ekert A 2022
  {\em Classical and Quantum Gravity\/} {\bf 40} 025013

\bibitem{bliokh2016transverse}
Bliokh K~Y 2016 Transverse spin and momentum in structured light: quantum spin
  hall effect and transverse optical force {\em 2016 URSI International
  Symposium on Electromagnetic Theory (EMTS)\/} (IEEE) pp 345--348

\bibitem{kholmetskii2007measurement}
Kholmetskii A~L, Missevitch O~V and Smirnov-Rueda R 2007 {\em Journal of
  Applied Physics\/} {\bf 102} 013529

\bibitem{dumont2020relativistic}
Dumont R~S, Rivlin T and Pollak E 2020 {\em New Journal of Physics\/} {\bf 22}
  093060

\bibitem{withayachumnankul2010systemized}
Withayachumnankul W, Fischer B~M, Ferguson B, Davis B~R and Abbott D 2010 {\em
  Proceedings of the IEEE\/} {\bf 98} 1775--1786

\bibitem{grigera2010dirac}
Grigera S 2010 {\em Bulletin of the American Physical Society\/} {\bf 55}
  X3.003

\bibitem{goldstein2002}
Goldstein H, Poole C and Safko J 2001 {\em Classical Mechanics\/} (San
  Francisco: Addison-Wesley)

\bibitem{lehmkuhl2014einstein}
Lehmkuhl D 2014 {\em Studies in History and Philosophy of Science Part B:
  Studies in History and Philosophy of Modern Physics\/} {\bf 46} 316--326

\bibitem{Oosten2000}
Van~Oosten A 2000 {\em The European Physical Journal D\/} {\bf 8} 9--12

\bibitem{brown2006minkowski}
Brown H~R and Pooley O 2006 {\em Philosophy and Foundations of Physics\/} {\bf
  1} 67--89

\bibitem{Cole1980comments}
Cole E~A~B 1980 {\em Physics Letters A\/} {\bf 76} 371--372

\bibitem{dorling1970dimensionality}
Dorling J 1970 {\em American Journal of Physics\/} {\bf 38} 539--540

\end{thebibliography}

\end{document}